\documentclass[aps,prl,article]{revtex4}

\usepackage{graphicx}
\usepackage{amsmath}
\usepackage{color}
\usepackage{natbib}
\usepackage{dcolumn}
\usepackage{bm}

\begin{document}

\title{Markovian Memory Embedded in Two-State Natural Processes}

\author{Fotini Pallikari}

\affiliation{University of Athens, Faculty of Physics, Department
of Solid State Physics, Panepistimiopolis Zografou, 15784 Athens,
Greece} \email{Electronic Mail: fpallik@phys.uoa.gr}

\author{Nikitas Papasimakis}
\affiliation{Optoelectronics Research Centre, University of
Southampton, Southampton SO17 1BJ, United Kingdom}

\begin{abstract}Markovian type of memory is considered as an inseparable ingredient in a variety of natural two-state processes within a vast 
range of interdisciplinary fields. The Markovian memory embedded in a binary system is shaping its evolution on the basis of its current state. In 
doing so, this type of memory introduces either clustering or dispersion of binary states.  The consequence is directly observed in the 
lengthening or shortening of the runs of the same binary state and also in the way the proportion of a state within a sequence of state measurements 
scatters about its true average. In the presence of clustering, this scatter will become broader.  Conversely, it will become narrower when dispersion 
of states is present. Both trends are directly quantifiable through the Markovian self-transition probabilities. It is shown that the Markovian 
memory can even imitate the evolution of a random process, regarding the long-term behavior of the frequencies of its binary states. This 
situation occurs when the associated binary state self-transition probabilities are balanced.  To exemplify the behavior of Markovian memory, two 
natural processes are selected from diverse scientific disciplines, belonging to a wide range of systems classified as two-state systems. The first 
example is studying the preferences of nonhuman troglodytes regarding handedness. The Markovian model in this case assesses the extent of 
influence two contiguous individuals may have on each other. The other example studies the hindering of the quantum state transitions that rapid 
state measurements introduce, known as the Quantum Zeno effect (QZE). Based on the current mathematical methodology, simulations of the experimentally 
observed clustering of states allowed for the estimation of the two self-transition probabilities in this quantum system. Through these, one can 
appreciate how the particular hindering of the evolution of a quantum state may have originated.  Namely, through a quantifiable degree of 
preference for the same binary state combined with a quantifiable degree of avoidance of the rival state.  The aim of this work is to illustrate 
as merits of the current mathematical approach, its wide range applicability and its potential to provide a variety of information regarding the 
dynamics of the studied process.\end{abstract}


\maketitle

\section{Introduction}

Markov sources provide phenomenological representations of a wide
range of natural processes. In a Markov chain involving state
measurements, the system under observation experiences state
transitions according to specified probabilities. A specific class
of Markov sources are the two-state systems, frequently employed
in physics on a multitude of occasions such as to represent spins
in the Ising model, or the outcome of particle collisions in the
Galton board binomial experiment\cite{Galton}. Binary systems can
also represent sequences of quantum measurements, as in the case
of the quantum Zeno effect (QZE), where the natural evolution of the
two-level atomic system is hindered by rapid observations
\cite{itanoPRA1990,BalzerOptComm2003}. Moreover, binary systems
are often encountered in electronic solid-state diffusion
\cite{hooverPLA88}, turbulent flow dynamics
\cite{chepPRL2001,LuePRE1993,fractal2006}, bistable liquid crystal
displays \cite{lagardePRE2003,xieJJAppPhys1998,forgerAdvMat2000},
 Josephson junctions \cite{baronePRL2004}, or flip-flop electrical
circuits in computing machines, which fit well as types
of two-state systems \cite{MoserIBMJ1961}. Binary Markov
processes can, therefore, describe adequately a whole host of
physical micro- as well as macro-systems
\cite{ritortJStatMech2004}.

Yet, two-state systems are not only pertinent in physics and
computing, but also highly relevant to probabilistic systems
commonly studied within the medical and social sciences. It is
customary, for instance, in medicine \cite{craenAmJEpidem2005},
psychiatry \cite{alemanAmJPsych1999}, or anthropology
\cite{palmerAmJPhysAnthr2002}, to test a hypothesis against two
alternatives. In medicine, the effect of a drug treatment is
decided from the proportion of successes (against failures) across
many independent studies. While the degree of reliability of the
results depends on the trustworthiness of the database, specific
tests are designed to determine whether the database is free from
biases. One such mainly visual test is the construction of a
scatter plot of independent results that constitute the database
\cite{alderson2004}. If the distribution of a large number of data
on the scatter plot appears asymmetric, it implies that the
database is biased by, as an example, publication biases and thus
rendered inappropriate source for reliable statistical inferences
regarding the studied effect. Less emphasis has been sited so far
on the scatter plot's breadth and the information that it can provide
concerning the dynamics of the underlying mechanism. The relationship
between them, however, can be exemplified on the basis of two-state
Markov processes, as will be shown in this work.

Markovian memory is introduced by the requirement that each new state
depends on previous states some n steps back, so that past
probabilities determine the future ones \cite{waner2004,reid1960}.
As it is discussed here, the dynamics of two-state systems fall
under two specific categories of Markovian memory, which either
introduce a clustering or a dispersion of binary states.

The purpose for the application of the Markov memory approach is to
study the first-order interaction dynamics in a variety of two-state
systems.  In doing so, it can estimate the conditional likelihood for
the system to change state or remain at the same state. It also estimates 
the consequent lengthening or shortening of runs of the same state in a 
sequence of state measurements.

A brief description of the two-state Markov process is offered in
section 2. The statistics of Markov chains in relation to the
clustering of states (lengthening of runs) and dispersion of states 
(shortening of runs) is discussed in section 3.
Two applications of the mathematical approach implicating Markovian
memory, in physics (QZE) and anthropology (Handedness) are
presented in section 4. A more detailed analysis of the present
mathematical treatment is provided in the appendix.

\section{First-order, two-state Markov process}

A first-order, two-state Markov process is driven by four
transition probabilities, $p_{ij}$, $i,j=1,2$. The sequence of measurements
of the binary state, $x_n$, represents the occurrence of an event
(state A), otherwise exemplified by $x_n=1$, or the failure of its
occurrence (state B) corresponding to $x_n=0$. The initial absolute
probabilities of finding the system at either state A or B are
$p_1$ and $1-p_1$, respectively. Once the
system is at state A, the conditional probability that it remains
at the same state after a single measurement is $p_{11}=p$,
whereas the conditional probability, $p_{21}$ to make a transition
to B will be $1-p$. In a similar fashion, probabilities $p_{22}=q$
and $p_{12}=1-q$ are assigned to transitions from state
B. In both cases, the self-transition probabilities satisfy the inequality
$0<p,q<1$. Whereas $p,q=0.5$ underlines random variability of
state measurements, the range of probabilities $0.5<p,q<1$ and $0<p,q<0.5$
introduce persistence of the same state and anti-persistence, respectively.
The latter case implies an increased probability to avoid transitions
to the same state over a sequence of measurements. We shall next
discuss how the expected frequency of the one state of the Markov process
(A), in a sequence of $n$ consecutive measurements, depends on the self-transition
probabilities $p$ and $q$.

The average frequency, $\bar{p_n}$, of occurrence of
state A in a Markov chain of $n$ steps is \cite{Mises}

\begin{equation}\label{eq1}
\bar{p_n} = \wp + \frac{p_1-\wp}{n} \cdot \frac{1-a^n}{1-a}
\end{equation}

The parameter $a\neq 1$ is

\begin{equation}\label{eq2}
a=p+q-1
\end{equation}

After a large enough number of state measurements, $n$, (Markov transitions)
the frequencies of observed states A and B become $\wp$ and $1-\wp$, respectively,
at any value of parameter $a$, $a\neq 1$.

\begin{equation}
 \wp={\lim_{n\rightarrow\infty} }p_{n}=\frac{1-q}{2-(p+q)}
\nonumber \end{equation}

\begin{equation}
 1-\wp=\frac{1-p}{2-(p+q)}\label{eq3}
\end{equation}

If  $p=q$, the frequencies of observed states A and B become $\wp=0.5$
and $1-\wp=0.5$. This result holds true not only in the absence of memory when $p=q=0.5$, but most
importantly, when $p=q \neq 0.5$.  This is a curious condition which
turns a Markov process with memory into a random process, as far as
the long-term state frequencies $\wp$ and $1-\wp$ are concerned.

Even in such an odd situation, the non-randomness of the Markov
process is directly observed through the variance of the binary
state in the sequences. The standard deviation of the expected
proportion of binary state A, $\bar{p_n}$, is estimated to be \cite{Mises}

\begin{equation}\label{eq4}
\sigma=\sqrt{\frac{\wp(1-\wp)}{n}}\cdot\sqrt{\frac{p+q}{2-(p+q)}}
\end{equation}

Relation (4) is also written $\sigma=\sigma_{0}\cdot\nu$, where
$\sigma_{0}=\sqrt{\wp(1-\wp)/n}$, is the standard deviation of
outcomes of a memory-free Markov process and it indicates that the
variance of $\bar{p_n}$ can be modulated by a factor $\nu^2\neq 1$
introduced by the Markov self-transition probabilities $p$ and $q$.
Assuming $p$ and $q\neq 0.5$ the factor which modulates the variance is

\begin{equation}\label{eq5}
\nu^2=\frac{p+q}{2-(p+q)}
\end{equation}

A large variety of natural processes can be represented as Markov
processes.  In such cases the characteristic parameters $\wp$ and
$\nu^2$ can be assigned accordingly. These provide insights into the
process dynamics on first neighbor level.  Often a process is
studied through its statistical behavior with the help of
meta-analyses. In such approaches, the two characteristic parameters, $\wp$ and
$\nu^2$, can be easily estimated through the so-called scatter
plots.  The application of the Markov process on such statistical
ensembles can provide useful information on the investigated
process as will be shown next.

\section{Scatter plot of Markovian binary states}

The combination of results from independent studies of a
phenomenon constitutes the so-called meta-analysis. The accuracy
of the result of each individual study depends proportionally on
the size, $n$, of the study, since the associated error is
inversely proportional to the square root of the standard
deviation.

The scatter plot, $n=f(\bar{p_{n}})$, where the size of studies,
$n$, is plotted against the associated proportion of binary state,
$\bar{p_{n}}$, will be shaped like an inverted funnel \cite{note1}
centered at $\wp$, the single true average, or in other words the
value to which the averages $\bar{p_{n}}$ converge. This is due to
the fact, as mentioned above, that the estimate of the underlying
effect becomes more accurate as the sample size of component studies
increases. Scatter plots can thus provide useful information not only on the
magnitude, $\wp$, of an investigated effect, but also about the
dynamics of the mechanism involved \cite{note2}.

The frequency of state A, $\bar{p_n}$, in a sequence of $n$ individual
measurements of a Markov process will range within a confidence interval.
The $95\%$ of them on the scatter plot are expected to be enclosed by the
confidence interval, represented by the two funnel-shaped red curves
in Fig. 1, $\bar{p_n}=f(n)$ or $n=f(\bar{p_n})$

\begin{equation}\label{eq6}
\bar{p_n}=\wp \pm 1.96\cdot\sqrt{\frac{\wp(1-\wp)}{n}}\cdot\nu
\end{equation}

and

\begin{equation}\label{eq7}
n=3.84\cdot\frac{\wp(1-\wp)\cdot\nu^2}{(\bar{p_n}-\wp)^2}
\end{equation}

\begin{figure}[h] \label{fig1}
\includegraphics[width=.65\textwidth]
{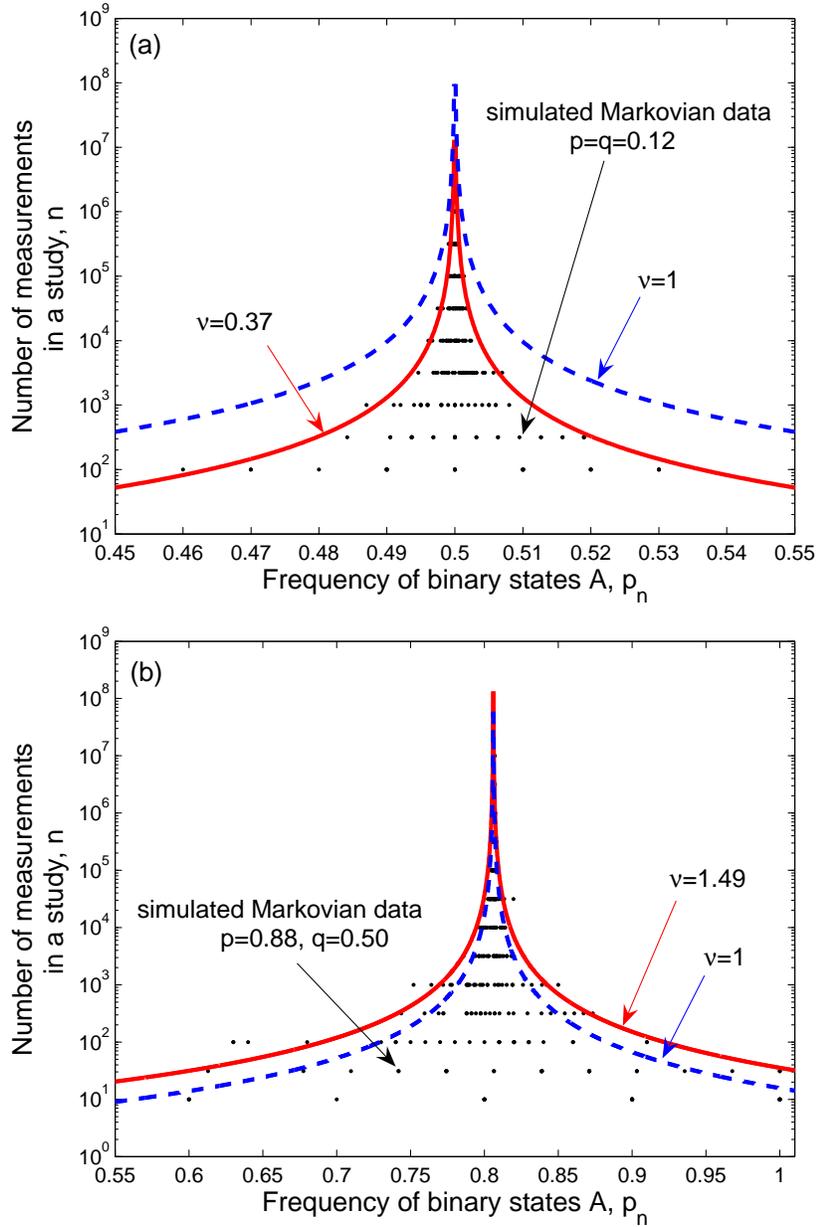}\caption{Scatter plot of Markovian data.  Dots:
computer simulated data, Lines: $95\%$ confidence interval. 
(a) Solid lines: Symmetric Markovian
process, $p=q=0.12$, $\nu=0.37$ and  $\wp=0.5$.  Probabilities
$p,q,<0.5$ introduce anti-persistence and $\nu<1$ introduces
dispersion of states and narrowing of variance. Dotted lines:
memory-free binary process, $\nu=1$.  (b) Solid lines: Asymmetric
Markovian process, $p=0.88$ and $q=0.5$, $\nu=1.49$, $\wp=0.81$.  The
persistence introduced by the self-transition probability, $p$, of
state-A, introduces broadening of the variance ($\nu>1$), according
to relation (4) and as compared to a Markov process with  $\wp=0.81$
and  $\nu=1$.}
\end{figure}

Computer-simulated Markovian data sequences of size $n$ were
generated and plotted against the frequency of binary state A,
$p_n$, in each sequence.  Fig. 1 illustrates two such examples;
(a) one of a symmetric $(p=q)$ Markov process
exhibiting anti-persistence, $p,q<0.5$, and (b) of an asymmetric
$(p\neq q)$ Markovian process with $p=0.88$, $q=0.5$ exhibiting
one-sided persistence.  In the symmetrical case, $p=q$, the
correlation factor $C_m$ between the two states in the Markov
chain

\begin{equation}\label{eq8}
C_m=\lim_{N\to\infty}\frac{1}{N}\sum_{n=1}^N s_n s_{n+m}
\end{equation}

becomes the assemble expectation value of
the correlation between first-order neighbors \cite{schroeder2001}

\begin{equation}\label{eq9}
C_1=<s_n s_{n+1}>=2p-1
\end{equation}

When $p=q=0.88$ the strong, positive correlation of neighbors
$(C1=+76\%)$ introduces persistence and therefore clustering of
states, while the condition $p=q=0.12$ introduces strong negative
correlation, anti-persistence and dispersion of states $(C1=-76\%)$.
The graphical representation of Fig. 1 confirms that the simulated
Markovian data obey well the statistical estimations: the $95\%$ of
them were found enclosed under the two $95\%$ confidence interval
curves, Eq. \eqref{eq7}, suggesting that this statistical tool
applied on scatter plots of meta-analyses is relatively reliable.

The modulated variance, as described above, indicates absence of
random variability in independent measurements of an effect
\cite{breslinPRA1997}. This type of observed irregularity within a
meta-analysis has been referred to by the term "statistical
heterogeneity" \cite{walkerAmJPubH1988}. There are numerous
reasons possible behind statistical heterogeneity in
a database. We consider here the two types that generate either
dispersion or clustering of Markovian states.

Looking closer at the shape of the plot we notice that, according to
Eqs. \eqref{eq4} and \eqref{eq5}, when $\nu>1$ its scatter becomes
broader at all sample sizes $n$, as compared that of a memory-free process
$(\nu=1)$. Similarly, the variance will be narrowed in the case
where $p+q<1$, (i.e. $\nu<1$). The broadening of a scatter plot is
the direct consequence of the persistence of a binary state and the
occurrence of longer-than-usual runs in the data sequences due to
state clustering, as Fig. 2 clearly illustrates.

\begin{figure}[h] \label{fig2}
\includegraphics[width=.65\textwidth]
{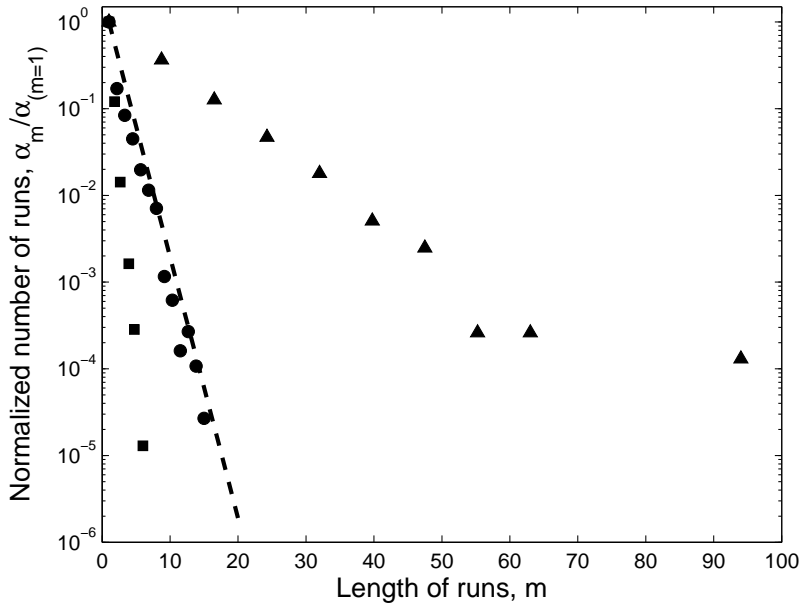}\caption{Normalized length of run (averaged over 10
sequences) in a 10,000-unit long computer generated Markovian
sequence.  (a) \textit{Circles}, $p=q=0.50$, memory-free Markovian
process; (b) \textit{triangles}, $p=q=0.88$, clustering of states;
(c) \textit{squares}, $p=q=0.12$, dispersion of states;  (d)
\textit{dotted line}: theoretical expectation for memory-free
processes, Eq. \eqref{eq10}.}\end{figure}

It is expected that, when $p=q=0.5$, the number of runs having
length $m$, $a_m$, i.e. sequences of the same binary state
occurring at frequency $\bar{p_n}$ in a sequence of total length
$n$, will be \cite{Mises}

\begin{equation}\label{eq10}
a_m=(n-m-1)\cdot[\bar{p_n}^2(1-\bar{p_n})^m+(1-\bar{p_n})^2\bar{p_n}^m]
\end{equation}

The variance $\sigma$ of $a_m$ for long enough sequences is:
$\sigma=a_m$. In Fig. 2 the normalized average number of runs (over
$10$ computer-simulated symmetric Markov sequences, each of length
$n=10.000$), is plotted against the length of run, $m$. When the
Markovian memory ($p=q=0.88$) introduces persistence and clustering
of the same binary state, the average length of the
computer-simulated runs is considerably longer than that of the
memory-free case ($p=q=0.5$). On the contrary, in the
anti-persistent case $(p=q=0.12)$, the average length of runs is
shorter than that of the memory-free sequences.  This result
supports that, in the latter condition, the states tend to disperse.
The simulated Markovian memory-free sequences, on the other hand,
behave according to Eq. \eqref{eq10}, as expected \cite{note3} by
theory. Estimating the self-transition probabilities ($p,q$) and the
variance factor $\nu^2$ provides a quantifiable description of the
Markovian dynamics in natural processes within a diversity of
scientific disciplines.

In the first example that follows, the above analysis will be directly 
applied on scatter plots referring to anthropological data. In the 
second example the analysis is applied on data recording two-state quantum 
transitions (QZE), to directly estimate the lengthening and shortening 
of runs of the same binary state. Then through simulations, based on 
the mathematical formulation developed here, it goes on to estimate the Markovian 
self-transition probabilities. From these, information regarding 
the transitions dynamics will be drawn.

\section{Examples of the Markov memory model in natural processes}

\subsection{Handedness as a Markov process}

Often in scientific disciplines such as medicine, anthropology,
sociology, or psychology the evaluation of evidence regarding an
effect makes use of scatter plots.  On one such occasion within
anthropology, a meta-analysis, that merged the results of 32
studies, was performed to investigate the strength and consistency
of right-handedness in nonhuman primates
\cite{palmerAmJPhysAnthr2002}, Fig. 3.  This application refers to
an even blend of primates raised in the wild as well as in
captivity. The constructed scatter plot in this study was intended to shed light
on questions regarding the consistency of the data variability with
normal sampling variation and the nature of biases in the reports
of statistically significant handedness. One could, however, get
additional information from these scatter plots, with regard to
the internal dynamics of the processes involved that
lead to handedness preference.

Anthropologists discovered that chimpanzees develop cultures in
the same way as humans do. According to Whiten et al.
\cite{whitenNat1999}, handedness in nonhuman primates tends to be
lateralized.  Yet, unlike humans where there is only $10\%$ use
of left hand, this proportion in nonhuman primates in the wild
reaches $50\%$.  Chimps in captivity, however, exhibit a weak
preference for the use of right hand, which is believed to result
from the human influence. These observations are in agreement with
results of the present Markov analysis.

We assume that the interactions among the nonhuman primates that
influence handedness are driven by a Markovian binary process, where
preference for the right and left hand is represented by the
self-transition probabilities, $p_{11}=p$ and $p_{22}=~q$,
respectively. Surely there exist multiple interactions among the
individuals in a group.  Yet, as it is a common practice in many
scientific disciplines, interactions are often reduced to first
neighbors, as long as this simplification does not distort their
true representation.  In Fig. 3 the number, n, of significantly-handed
individuals (pan troglodytes) was plotted against the proportion,
$\bar{p_n}$, of them who were right-handed. The confidence interval
curves according to Eq. \eqref{eq7} that best envelope the $95\%$ of
the experimental data are drawn by adjusting the Markov memory
parameters and are given by

\begin{equation}\label{eq11}
n=\frac{1.24}{(\bar{p_n}-0.58)^2}
\end{equation}

The confidence interval representing data of random variability
$(\nu =1)$, which implies an equal preference for right and left
hand use, is also marked on the graph for comparison.  It is
clearly not fitting adequately the totality of these data, apart
from these referring to individuals living in the wild.

The Markov analysis rendered the parameters, $p=0.64$, $q=0.50$,
$\nu=1.15$, $\wp=58\%$ for the individuals in captivity and
$p=q=0.50$, $\nu=1$,  $\wp=50\%$ for those in the wild (dark circles).
The accuracy of the three parameters above is limited to $1\%$ and
so the parameters are rounded up to the second decimal digit \cite{note4}.
It is a known fact that the members of a group positively influence
each other with regard to handedness. Therefore, the estimated self-transition
probabilities for either the use of the right or the left hand should
be $\geq0.5$.  That fact together with the condition that the $95\%$ of
the data points should be enveloped by the confidence interval
curves enables a relatively accurate adjustment of the two
parameters,  $p$ and $q$, within the limitations of the size of
this database.

\begin{figure}[h] \label{fig3}
\includegraphics[width=.65\textwidth]
{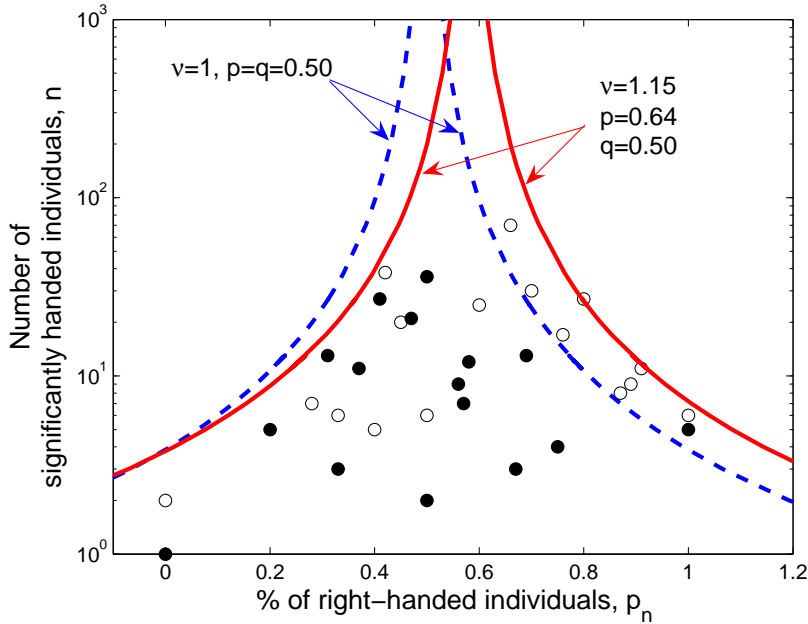}\caption{Scatter plot of a meta-analysis investigating
preference in handedness among nonhuman primates
\cite{palmerAmJPhysAnthr2002}. Open circles: captive, closed
circles: individuals in the wild.  The $95\%$ confidence interval
curves are also plotted (solid curves), Eq. \eqref{eq7}, suggesting
that $\wp=0.58$, $p=0.64$ and $q=0.50$, $\nu=1.15$ for the totality 
of the data. Dotted lines: memory-free and symmetric Markov process 
($p=q=0.50$, $\wp=0.5$, $\nu=1$) best representing individuals in the 
wild. See text for details.}\end{figure}

The current results can be understood as follows. They first imply
that the studied individuals have a tendency to use both left and
right hands equally, unless they are in captivity (open circles).
In that latter case, the human influence modifies their handedness
habits:  Therefore, the initial $50\%$ preference for the use of
the right hand rises to a $58\%$ ($\wp=0.58$). This conclusion
represents only part of the information available by the Markov
memory analysis. As it confirms the independent observations in the
reference study, one can be relatively confident about the effectiveness,
reliability and applicability of the Markov approach used.

Apart from the static average value above, the current Markov analysis
provides extra information towards the understanding of the
chimpanzees' interactions. The collective evaluation of independent
studies in a meta-analysis presupposes that they are all measures
of one the same phenomenon within samples of different size. We can,
therefore, assume that the size of a group of the individuals in
captivity can increase for a number of reasons (that do not concern
the scope of this paper). Every newly added member would take after
the handedness habit of the one close to them, whether this is a mother,
a partner etc, in the following likely scheme.  In this representation
of handedness habits of individuals in captivity by the first-order Markov 
memory model, there will be a $64\%$ probability that an added member 
of the group will prefer to use their right hand, as the individual 
close to them does. Also that $36\%$ of them will prefer to use their 
left hand instead, if they are close to a right-handed individual.  The use 
of the left hand appears not predisposed by human presence, however. There will 
be a balanced $50\%$ probability to use either their left or right hand 
if the individual close to them is a left-handed individual.

The merit of the current Markov analysis, when applied on scatter plots,
lies on the fact that it provides in one graph more information than just
the static parameters directly available from other qualified
statistical methods. Additionally, it offers a quantitative description
of the interactions between the studied system units.  In that sense, the
current analysis sheds additional light on the dynamics of the system
under study which otherwise constitute complicated processes. We showed
here that the Markov analysis widens the scope of applicability of scatter
plots, from mere graphic examinations of the presence of biases in data
bases to effective tools for the understanding of the interactions among 
the group units.

\subsection{The quantum Zeno effect}

The second example treats a quantum two-state system exhibiting the 
so-called quantum Zeno effect (QZE) \cite{itanoPRA1990,BalzerOptComm2003}, 
that has taken its name from the famous Zeno paradox. In the QZE the 
natural state transitions can be impeded by fast repetitive state-monitoring 
measurements. Therefore, a series of state measurements will also observe 
the persistence of this preferable state from which transitions have been 
hindered.  Longer runs of that state will be recorded in a series of 
measurements, as Fig. 2 illustrates.  Therefore, the persisting state 
will appear to cluster while the state that is avoided will appear to 
disperse in the sequence of state measurements. The clustering is marked 
by an increased frequency of runs of the same state (triangles in Fig. 2) 
as compared to the memory-free unhindered situation (circles and dotted 
line in Fig. 2).  As the current Markov memory model mathematically handles 
this clustering and dispersing behavior of the two-state system, the QZE 
represents an attractive candidate for its application.

Specific details about the experimental settings and the generation of
quantum states by Balzer et al that can be found directly in the source
paper are not pertinent to the current analysis.  It suffices to mention
that in this quantum system the transitions between the two-states are
driven by the application of an appropriate field.  As the system evolves
in time driven between the two states, it is possible to observe at which
state the system is at a certain time by the application of a second field,
the probe field.  The probe field monitors the state of the system.  Emission
of scattered light indicates that the system is at the lower state-1, called
'on' state.  Absence of scattered light indicates that the system is at the
upper state-2 called 'off' state.  It was observed that under certain
experimental settings introducing fast repeated state measurements, the evolution
of the quantum system was hindered.  The suppression of transitions is 
observed by a clustering of the state that persists, or by the dispersion of the
second state, in a long sequence of state measurements. It was documented
in the source paper by a graph of the normalized frequency of "uninterrupted
sequences" of state 'on' and state 'off', against the length of the corresponding
sequences.  The degree to which the probing process interacted with the
drive field to impede the system's quantum evolution was thus established.

The Markov memory approach developed in this paper goes a step
further to estimate the conditional probabilities that determine
the readiness of the two-state system to make a transition, or 
remain at the same state with each probing measurement.  The 
frequency of uninterrupted sequences (runs) of states 'on' and 
'off' were simulated, by the selection of appropriate self-transition 
probabilities, and fitted on the experimental data.  The associated 
first-order Markov self-transition probabilities, $p_{11}$ and $p_{22}$ 
thus estimated at each of the three experimental settings employed, 
provide an insight into the state transition dynamics.

\begin{figure}[h] \label{fig4}
\includegraphics[width=.65\textwidth]
{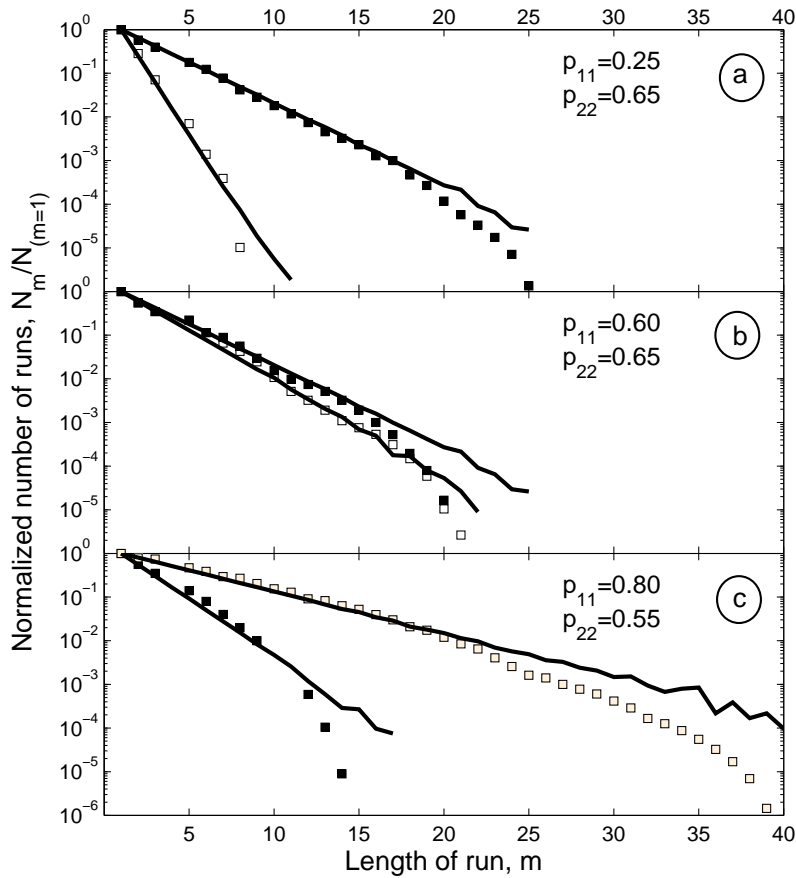}\caption{Clustering and dispersion of Markovian states in
the quantum Zeno effect.  Experimental data for three different
detuning settings \cite{BalzerOptComm2003,note6}. Open squares: 'on' events, state-1.
Dark squares: 'off' events, state-2.  Lines: computer simulated runs
of a two-state Markov memory process, modulated by the
self-transition probabilities, $p_{11}$ and $p_{22}$.  See text for
details.}\end{figure}

In the first experimental setting, the system starts from state-2 at conditions
favoring its persistence.  The estimated self-transition Markov probabilities
were $p_{22}=0.65$ and $p_{11}=0.25$, or equivalently $p_{21}=0.75$. The latter
probability implies that once the system is at state-1 it is very eager to make a
transition back to state-2.  On the other hand, once the system is at state-2 it
prefers to remain at it by a $65\%$ probability with every consequative measurement. 
The system therefore, was lead to persistence of the 'off' state-2 due to a 
combination of two factors: its preference to remain at this state combined with 
the most effective $75\%$ preference to avoid the competing 'on' state-1. This 
analysis shows therefore that a quantum transition is impeded not only because 
the experimental conditions favor that state, but mainly because they obstruct 
the rival state.  Longer runs of state-2 will be observed in that case, Fig. 4a.

There were further two more experimental settings employed in this experiment.  The
first one, fig. 4c, favors the persistence of state-1. The application of the Markov memory analysis
on these data has turned the values $p_{11}=0.80$ and $p_{22}=0.55$. These imply
that once the system is at state-1 transitions from it are hindered, because it
exhibits a strong $80\%$ preference to remain at this state combined with a small
$45\%$ tendency to avoid state-2. Unlike in the previous experimental setting, here
it is mainly the strong preference for state-1 that hinders transitions from it.

Finally, the third experimental setting brings the quantum system somewhere in between
the previous two behaviors, fig. 4b. The Markov memory analysis has rendered the values
$p_{11}=0.60$ and $p_{22}=0.65$. In this case there is an equal, yet relatively weak,
$60-65\%$ preference for the system to remain at either state. The evolution of the
system from both states is moderately hindered, but the experimental conditions level 
the competition between the two rival states. Thus, the system shows evidence of similarly 
administered preference, where state-2 is temporarily and mildly winning.

This example shows how the QZE can alternatively be accounted for as
a process driven by a Markovian type of memory.  In that procedure,
an insight regarding the inner dynamics of the observed effect is 
obtained. The agreement between simulated and experimentally observed 
QZE data is very good, especially for frequencies up to about $10^{-4}$.

\section{Summary}
In this work it is shown how the natural evolution of two-state processes is shaped by the presence of Markovian memory in them.  The presented 
mathematical formulation provides an alternative, quantitative as well as qualitative, description of a wide interdisciplinary range of processes, 
classified as two-state systems.  The presence of Markov memory, \cite{note7} will enhance either the clustering or the dispersion of these binary 
states. When independent studies of a phenomenon are combined and graphically presented in scatter plots, for instance, the clustering of states is 
recognized by a broadening of the associated scatter plot.  Conversely, the dispersion of states is marked by a narrowing of the scatter plot's 
breadth.  These changes occur for the reason that the Markovian memory has affected the length of runs of the same state in the two-state system.  
Two examples are invoked to exemplify the applicability of the Markov memory approach and the kind of prospective information that can be gained 
from it.  The first one is a two-state system taken from anthropology, while the second one is taken from quantum theory.
 
In the first example, independent studies testing handedness in a mixture of cultures of chimpanzees were shown to exhibit Markov memory with 
regard to how strongly-related individuals may influence each other. Individuals raised in the wild exhibit equal preference for the use of both 
hands, regardless of what is the preference of their closest relatives. The presence of humans influences this attitude as far as the use of the 
right hand is concerned.  There was, therefore, a mildly increased preference for the use of the right hand assessed to a $64\%$ probability, that 
the closely related individuals exhibit the same preference.  There was no similar effect observed in the use of the left hand, 
however.  It appears that a new member of the group, closely associating with a left-handed individual will exhibit an independent handedness 
preference.  Overall, a proportion of $58\%$ of individuals across the mixture of studied groups, both in captivity and in the 
wild, prefer to use their right hand. These estimated tendencies in handedness, influenced by presence of humans, agree with independent 
anthropological studies, conveying a degree of confidence to them.

The current Markovian mathematical formulation was also applied on the experimental observation of the quantum Zeno effect (QZE) in a two-state 
system. The experimental evidence in this example exhibits a combination of characteristic clustering and dispersion of the binary states, across
the three experimental conditions employed by the source paper of Balzer et al.  Through simulations of the length of runs in each of the three 
experimental conditions, it was possible to observe that a synchronous clustering of the one state and dispersion of the other can be responsible for 
the characteristic hindering of state transitions. The probabilities that determine the likelihood for the system to make one transition to the other 
state, or remain at the same binary state, were thus estimated.  These probabilities should not be confused with those estimating the frequency of 
each state in a long sequence of measurements.  It was, thus, concluded that the hindering of a binary state in the quantum Zeno effect may be 
effected not so much by the $65\%$ preference for state-2 through two consecutive measurements, as through the synchronous avoidance of the rival 
binary state-1, by a $75\%$ probability that the rival state will precede the preferred state.  Similarly, the hindering of the evolution of state-1 
can be effected by a strong $80\%$ preference for that state, winning over a weak preference $55\%$ for the rival state.  

The Markov memory analysis of two-state processes presented here treats complicated processes in terms of their first-order, first-neighbor 
interactions.  This simplification is an initial step towards the understanding of rather complex processes that can describe the system and where 
comparisons are possible, are in agreement with experimental evidence. Our understanding of nature builds in steps of gradual complication. 
Occasionally, simplified versions of reality, rather than the more complicated or unattainable ones are tried, as long as their exercise does not 
conflict with established experimental evidence.

\section{Appendix A}

Following the treatment of von Mises \cite{Mises} we assume that
$|p+q-1|\neq1$ excluding the cases $p=q=1$ and $p=q=0$, as they
present no interest but $0<p,q<1$. The following recursion formula
holds for the state probabilities

\begin{equation}\label{eq16}
p_i^{(n)}=\sum_{j=1}^2p_{ij}p_j^{(n-1)},~i=1,2;~n=1,2,...
\end{equation}

Equations \eqref{eq16} are equivalent to the iteration set up of the
following homogeneous equations

\begin{equation}\label{eq17}
-x_i+\sum_{j=1}^2p_{ij}x_j=0,~i=1,2
\end{equation}

The sum of transition probabilities
of each column is equal to $1$. Also the absolute probabilities at
every step of the system's evolution sum up to unity

\begin{equation}\label{eq18}
\sum_{j=1}^2p_{ij}=1
\end{equation}

\begin{equation}\label{eq19}
\sum_{j=1}^2p_{i}^{(n)}=\sum_{j=1}^2p_{i}^{(0)}=1
\end{equation}

 Since the 2x2 transition matrix having elements $p_{ij}$ $i,j=1,2$ is regular,
$\underline{p}^{(n)}$ tends to a unique fixed
  probability vector $\underline{p}^{(\infty)}$ that can be estimated
by solving the system of Eqs. \eqref{eq17}. These lead to two
non-zero solutions $u$ and $1-u$, to which the probabilities $p_n$
and $1-p_n$ converge after a large number of trials, $n$.  The two
roots of system (17) are the values $\lambda$ for which its
determinant $|P(\lambda)|$ vanishes

\begin{equation}\label{eq20}
 |P(\lambda)|=\begin{vmatrix}
p-\lambda & 1-q \\
1-p & q-\lambda
\end{vmatrix}
=0
\end{equation}

where $\lambda$ cannot be greater than $1$ in absolute value.  In
this case the two roots are

\begin{equation}
\lambda_1=p+q-1 \nonumber
\end{equation}
\begin{equation}\label{eq21}
\lambda_2=1
\end{equation}

The equations \eqref{eq16}, and \eqref{eq21}, can then be written as

\begin{equation}
u_1=p\cdot u_1+(1-q)\cdot u_2 \nonumber
\end{equation}
\begin{equation}\label{eq22}
u_2=(1-p)\cdot u_1+q\cdot u_2
\end{equation}

 yielding the two not uniquely determined solutions $u$ and $1-u$

\begin{equation}
u\equiv\wp=\lim_{n\to\infty}p_n=\frac{1-q}{2-(p+q)} \nonumber
\end{equation}

\begin{equation}\label{eq23}
1-u \equiv 1-\wp=\frac{1-p}{2-(p+q)}
\end{equation}

Equations \eqref{eq23} have an important consequence.  The
probabilities of finding the system at either binary state after n
trials converge to $\wp=50\%$, provided the two self-transition
probabilities are equal, $p=q$, and regardless if they are different
from $50\%$.  In other words, in the long run binary states (A and
B) will occur in the Markov chain at the same frequency as if no
memory was involved. From the point of view of long-run state
probabilities the Markov process will resemble a memoryless
Bernoulli case.  We shall next estimate the frequency of state A.

As stated in the text, the number $x_\nu=1$  is associated with the occurrence of
state A after a measurement and $x_\nu=0$ is associated with the
occurrence of state B.  The expectation value of $x_\nu$, i.e.
$p_\nu$ , will be the probability that the outcome of the
$\nu^{th}$ measurement will be the number $1$.  The expectation
value of all the '1' states present in a Markov sequence of n
trials will be then

\begin{equation}\label{eq24}
E[x_1+x_2+...+x_n]=p_1+p_2+...+p_n
\end{equation}

and the expected value of proportion of ones in the chain in will
be

\begin{equation}\label{eq25}
E[\frac{x}{n}]=\frac{p_1+p_2+...+p_n}{n}=\bar{p_n}
\end{equation}

The recursion formula (16) can be written
\begin{equation}\label{eq26}
p_n=\sum_{j=1}^2p_{1j}p_j^{(n-1)}=p_{11}p_1^{(n-1)}+p_{12}p_2^{(n-1)}=a\cdot
p_{n+1}+b
\end{equation}

where $a=p+q-1\neq1$, and $b=1-q$ since $0<p<1$ and $0<q<1$. Given
that the initial value of $p_n$ is $p_1^{(0)}=p_1$ , the recursion
formula (26) yields \cite{note8}

\begin{equation}\label{eq27}
p_n=a^(n-1)p_1+\underset{(n-1)
terms}{\underbrace{a^{n-2}+a^{n-3}+...+a+1}}\cdot b=
a^{n-1}p_1+\frac{1-a^{n-1}}{1-a}\cdot b
\end{equation}

or

\begin{equation}\label{eq28}
p_n=a^{n-1}{\huge [
}p_1-\frac{1-q}{2-(p+q)}+\frac{1-q}{2-(p+q)}=a^{n-1}[p_1-\wp]+\wp
\end{equation}

where $\wp$  is defined in (23). The average expected proportion of
state A, $p_n$, in the chain of $n$ trials would be written as
\cite{note9}

\begin{equation}\label{eq29}
\bar{p_n}=\frac{1}{n}\sum_{i=1}^np_i=\frac{1}{n}(p_1+p_2+..+p_n)=\wp+\frac{p_1-\wp}{n}\cdot\frac{1-a^n}{1-a}
\end{equation}

and

\begin{equation}\label{eq30}
\bar{p_n}\underset{n\rightarrow\infty}{\rightarrow}{\wp}
\end{equation}

The absolute probability of finding the system at state A after
$n$ measurements of its state is written $p_1^{(n)}=p_n$ and the
probability of finding it at state B is $p_2^{(n)}=1-p_n$.  This
is equivalent to saying that the $n$-th measurement of the
system's state has a probability $p_n$  of finding it at state A.
Since $p^{(n)}=p_{11}^{(n)}~ and ~ q^{(n)}=p_{12}^{(n)}$ the
recursion formula (16) applies to the transition probabilities
between states too

\begin{equation}
p^{(n)}=a\cdot p^{(n-1)}+b \nonumber
\end{equation}

\begin{equation}\label{eq31}
q^{(n)}=a\cdot  q^{(n-1)}+b
\end{equation}

The recursion formula then yields, since
$p_{11}^{(0)}=1~ and ~p_{12}^{(0)}=0$

\begin{equation}\label{eq32}
p^{(n)}=a^np^{(0)}+\underset{n~
terms}{\underbrace{[a^{n-1}+a^{n-2}+...+a+1]}}\cdot
b=a^n+\frac{1-a^n}{1-a}\cdot b=a^n\cdot[1-\wp]+\wp
\end{equation}

and

\begin{equation}\label{eq33}
q^{(n)}=a^nq^{(0)}+\underset{n~
terms}{\underbrace{[a^{n-1}+a^{n-2}+...+a+1]}}\cdot
b=a^n\cdot[1-\wp]
\end{equation}

If the system is at state A the probability that after $n$ steps,
where $n$ is very large, the measurement will yield again state A is
$\wp$: $p^{(n)}\underset{n\rightarrow\infty}{\rightarrow}\wp$ and
to yield state B is zero
$q^{(n)}\underset{n\rightarrow\infty}{\rightarrow}0$.

Von Mises has estimated the standard deviation of the mean value
$\wp$, in other words, the asymptotic value of the standard
deviation of the proportion of state A, $\bar{p_n}$, in the
sequence of $n$ trials as

\begin{equation}\label{eq34}
\sigma=\sqrt{\frac{\wp(1-\wp)}{n}}\cdot\sqrt{\frac{1+a}{1-a}}=\sigma_o\cdot\nu
\end{equation}

The proportion of state A in sequences of trials of
length $n$ of a Markov process with memory scatters about the asymptotic
value $\wp$ , while their scatter has been modulated with respect to the
memory-less case

\begin{equation}\label{eq35}
\sigma_o=\sqrt{\frac{\wp(1-\wp)}{n}}
\end{equation}

The modulating variance factor in (34)
  \begin{equation}\label{eq36}
\nu=\sqrt{\frac{1+a}{1-a}}
\end{equation}

takes values either above or below 1 depending on the self transition probabilities according to

\begin{equation}
\nu>1\Rightarrow p+q>0.5 \nonumber
\end{equation}

\begin{equation}\label{eq37}
\nu<1\Rightarrow p+q<0.5
\end{equation}


\begin{thebibliography}{34}
\bibitem{note1} Also termed the funnel plot due to its shape.
\bibitem{note2} Trustworthiness of statistical results is, however, limited by insufficiently small databases and the presences of biases.
\bibitem{note7} Determined by the fact that the outcome of a measurement depends on the previous measurement.
\bibitem{note8} The sum of the converging geometric series $(a<1)$ has $n-1$ terms.
\bibitem{note9} The sum of the converging geometric series has now $n$ terms.
\end{thebibliography}
\end{document}